\documentclass{aastex}
\usepackage{graphicx}

\usepackage{amssymb}
 
 \usepackage{natbib}

\begin{document}

\title{Evidence for condensed-phase methane enhancement over Xanadu on Titan}

\author{M. \'Ad\'amkovics and I. de Pater}
\affil{Astronomy Department, University of California,
    Berkeley, CA 94720}

\author{M. Hartung}
\affil{Gemini Observatory, La Serena, Chile}

\author{J. W. Barnes}
\affil{Department of Physics, University of Idaho, Moscow, ID 83844}

\maketitle

\begin{abstract}
We present evidence for condensed phase methane precipitation near Xanadu using
nine nights of observations from the SINFONI integral-field spectrograph at the Very Large Telescope and imaging analysis with empirical surface subtraction. Radiative transfer models are used to support the imaging technique by simulating the spectrometer datacubes and testing for variations in both the surface reflectivity spectrum and atmospheric opacity. We use the models and observations together to argue against artifacts that may arise in the image analysis. High phase angle observations from Cassini/VIMS are used to test against surface scattering artifacts that may be confused with sources of atmospheric opacity. Although changes in the surface reflectivity spectrum can reproduce observations from a particular viewing geometry on a given night, multiple observations are best modeled by condensed-phase methane opacity near the surface. These observations and modeling indicate that the condensed-phase methane opacity observed with this technique occurs predominantly near Xanadu and is most likely due to precipitation.
\end{abstract}

\section{Introduction}

A methane-based meteorological cycle involving clouds and rain has been suspected to occur on Titan since the time of Voyager, when methane was measured by IRIS \citep{Hanel1981}. Pioneering radiative transfer models of the atmosphere \citep{Toon1989, McKay1989} were used to interpret these spectra, predicting rain and patchy clouds from 10 -- 30\,km \citep{Toon1988}. Ground-based measurements at shorter wavelengths supported the presence of clouds and discovered their daily variation \citep{Griffith1998, Griffith2000}. Advances in imaging technologies lead to spatially-resolved observations of clouds near the south pole \citep{Brown2002,Roe2002b}, as well as a region of clouds confined to southern mid-latitudes \citep{Roe2005a}.
Since the arrival of Cassini it has been possible to resolve complex cloud structures \citep{Porco2005}, measure updraft velocities \citep{Griffith2005}, and identify polar clouds that may be composed of ethane \citep{Griffith2006}. 

Global scale precipitation has been suggested using the methane relative humidity profile measured by Huygens \citep{Tokano2006a}. While it was first predicted that  droplets would evaporate before reaching the surface \citep{Lorenz1993}, recent models suggest rain can fall to the ground \citep{Graves2008}. Fluvial channels indicate that a liquid was in contact with the surface at some time \citep{Tomasko2005}. 

Near-IR observations in two methane transmission windows (1.6 and 2.0\,$\mu$m) have been used with radiative transfer models to measure the column of condensed-phase methane on Titan \citep{Adamkovics2007}. These data revealed a global cloud of solid methane in the 25 - 35\,km altitude range, with an enhancement of opacity on the morning hemisphere that is consistent with methane drizzle. Recently, these results have been called into question using observations made at the Gemini-North telescope with the NIFS spectrometer \citep{Kim2008}. 

Here we present measurements over consecutive nights, together with radiative transfer modeling, to show that the signature of increased condensed-phase methane cannot be due to surface reflectivity artifacts as suggested by \citet{Kim2008}, and use high (Sun-target-observer) phase angle measurements to rule out phase angle artifacts. We account for the conclusions and analysis of \citet{Kim2008} by quantifying the errors that can arise in the surface subtraction due to noise and bias. Our observations indicate that while precipitation occurs predominantly near Xanadu, it is not solely a morning phenomenon as suggested earlier \citep{Adamkovics2007}.

\section{Observations}

Ground-based observations were performed between 2006 Dec 28 and 2007 Mar 11 UT with the adaptive-optics aided integral-field spectrometer, SINFONI \citep{Eisenhauer2003c}, at the Very Large Telescope (VLT). The spectrometer uses two sets of 32 stacked mirrors to divide the field of view (FOV) into 32 `slitlets', which are then optically arranged into a single synthetic long slit that is dispersed onto a cryogenically-cooled 2K $\times$ 2K Hawaii detector. Each of the 32 slitlets is imaged onto 64 $\times$ 2000 pixels on the detector. We used the $0.8'' \times 0.8''$ FOV, corresponding to a pixel scale of $0.0125'' \times 0.025$'', with the H+K grating covering 1.45 -- 2.45\,$\mu$m at a resolving power, $R = \lambda / \Delta\lambda$, of 2000 to 3400. 

Data-reduction follows the SINFONI pipeline \citep{Modigliani2007}. The pipeline includes a reference bad-pixel map, a check for pixels approaching a non-linear detector response, dark current correction, and flat fielding.
All of the spectra are distortion (curvature) corrected using standard flats and arc lamps frames, along with wavelength calibration from a reference arc line table. Spectra are interpolated onto a common wavelength grid before reconstruction of the datacube using the slitlet position and slitlet edge tables.

Mosaicking is performed using AO pointing keywords to overlap images to the nearest pixel and then taking the median value for overlapping pixels. Telluric correction is performed with the extracted spectra of standard stars. Flux calibration is performed by normalization to the disk-integrated stratospheric flux from Keck/OSIRIS observations \citep{Adamkovics2007}, and are reported in units of albedo, $I/F$. The observation dates and viewing geometries are given in Table 1.

The Visual and Infrared Mapping Spectrometer (VIMS) on the Cassini spacecraft uses a spectral mapping technique \citep{Brown2004} to assemble images at 352 wavelengths, covering 0.3 to 5.1\,$\mu$m. Channels near 1.6 and 2.0\,$\mu$m can be used for direct comparison with ground-based observations, and provide the opportunity for high phase angle viewing geometries that are not obtainable from Earth. We use VIMS data that are publicly available from the T10 flyby on 2006 January 15 UT. The data are reduced from observed data numbers to units of albedo, $I/F$, according to \citet{Barnes2007}. 

\section{Radiative Transfer Model}

In order to interpret the observed spectral datacubes, we create a forward model to simulate the flux that is reflected from Titan's atmosphere and surface. Our model incorporates well-established numerical solutions to the radiative transfer (RT) equation \citep{Griffith1991, Brown2002, Adamkovics2004, Adamkovics2007} that reproduce near-IR observations exceptionally well. 
Updates to the model presented in \citet{Adamkovics2007} include the treatment of the surface reflectivity; including the specification of a unique surface spectrum for Xanadu, averaging the surface albedo map over the footprint of a pixel in the model, and an explicit fit for converting the observed surface albedo map to a surface reflectivity map that is input to the model (described in Section 3.3).
Limitations in current analyses are due to input parameters  (e.g., CH$_{4}$ opacities, surface albedo spectra, and aerosol scattering) and not the numerical methodology or assumed geometry. A 2-stream approximation (at small phase angles) or discrete ordinates method (at high phase angles) is used to solve the radiative transfer equation for 16 pseudo-plane-parallel layers from 0 -- 200~km altitude \citep{Toon1989,McKay1989,Stamnes1988}. We correct for the curvature of the atmosphere with an established geometrical correction \citep{Tran2004}. We use the atmospheric temperature and pressure profiles measured {\em in situ} by Huygens/HASI \citep{Fulchignoni2005}, which are in excellent agreement (below 200~km) with the Voyager profiles \citep{Lellouch1989}. Starting with the {\em in situ} measurements of aerosol extinctions made by Hugyens/DISR \citep{Tomasko2005}, we vary the extinction at other locations on Titan to reproduce our observations of the global distributions of haze. The uncertainty of the CH$_{4}$ opacity, which is greatest for weak transitions, translates into an uncertainty in both the retrieved surface albedo and to a lesser extent the tropospheric aerosol extinction that is constrained at other wavelengths.

\subsection{Gas opacity}

The vertical profile of CH$_4$ mixing ratio measured by Huygens/GCMS \citep{Niemann2005} is applied globally and the absorption coefficients for CH$_4$ presented by \citet{Irwin2005} are used to calculate CH$_4$ opacities. Corrections to these coefficients have been published using the measurements made by Huygens/DISR \citep{Tomasko2008c}, however the corrections do not extend to 2$\,\mu$m and are not applied here. An alternative determination of the CH$_{4}$ opacity may be performed with line-by-line calculations that use the very high resolution, high sensitivity, CH$_{4}$ spectra of Brown \citep{Brown2005}. These data are currently available in the HITRAN database. However, the line strengths for these transitions are only applicable at room temperature, and quantum mechanical assignments of individual CH$_{4}$ lines are required before opacities can be calculated at the temperatures applicable to Titan. As described by \citet{Brown2005}, assigning these transitions requires considerable laboratory and computational effort.

Other sources of gas-phase opacity in the 2.0\,$\mu$m~band include low temperature collision-induced absorption by N$_2$ and H$_2$ \citep{McKellar1989}. The H$_2$-N$_2$ collision complex is a significant source of opacity near 2.1\,$\mu$m, which decreases the observed  slope of Titan's spectrum in the 2.0\,$\mu$m~band relative to the 1.5\,$\mu$m~band. Since the magnitude of the observed flux can be roughly mapped onto the altitude where most of the scattered photons originate --- that is, regions of weak atmospheric absorption are bright and probe the surface, whereas regions of high gas opacity are dim and probe the upper atmosphere --- the gradual variation in the 2.0\,$\mu$m~band extinction facilitates altitude discrimination relative to the 1.5\,$\mu$m~band. This collision-induced opacity is critically dependent on the mixing ratio of molecular hydrogen, which we take to be well mixed throughout the atmosphere at levels of $\sim$0.1\% \citep{Samuelson1997}. 

\subsection{Aerosol Haze Scattering}

Observations at IR wavelengths can be represented by models with an approximated scattering phase function (e.g., a one-term Henyey-Greenstein). When calculating the IR flux reflected from Titan at low phase angles there is, in general, a degeneracy between the aerosol scattering phase function and the extinction. For example, an increase in albedo can result either from a greater aerosol density or aerosols that are more backscattering. 

The DISR instrument on Huygens was used to measure the haze extinction profile from 0.5 -- 1.5\,$\mu$m \citep{Tomasko2005}, and our ground-based observations made during probe entry (at 1.5\,$\mu$m) use these measurements as a benchmark for the extinction near the latitude of the Huygens landing site, 10$^{\circ}$S. The cumulative DISR extinction is extrapolated to observations at 2\,$\mu$m \citep{Adamkovics2007}. Recently, a parameterization of aerosol scattering in three altitude regimes on Titan has been fit to the DISR data \citep{Tomasko2008a}. The authors propose a model of the wavelength dependence of the scattering phase function that is derived using the DISR data and scattering that is treated with the T-matrix method \citep{Mishchenko1996}. The net solar flux both in Titan's atmosphere and at the surface, calculated using the detailed optical properties and distribution of particles, is in good agreement with the values calculated by \citet{McKay1989} \citep{Tomasko2008b}.

We use a simplified parameterization of the phase function, following \citet{McKay1989}. The two-stream methodology we use incorporates the total flux that is scattered in the forward and backward directions. This method is not sensitive to further subtleties in the scattering phase function. For the 1.5 and 2.0\,$\mu$m windows, a one-parameter Henyey-Greenstein phase function has been adequate for interpreting the observations and has accurately predicted the heat balance in the atmosphere \citep{Tomasko2008b}.

The spatial variation in aerosol density is modeled using the method presented in \citet{Adamkovics2006}. The stratospheric aerosol extinction increases northward at a rate of 0.55\% per degree latitude from 45$^{\circ}$S up to $\sim$70$^{\circ}$N,  the highest latitudes observed here. The tropospheric aerosol extinction is uniform below 40\,km altitude.

\subsection{Surface reflectivity}

The surface reflectivity is modeled with the simplest representation that can reproduce both spatial and spectral variations in the observations. 
We start with a high-resolution cylindrically-projected map of the 2.018\,$\mu$m albedo \citep{Barnes2007} and re-project it onto a sphere viewed on the plane of the sky, at the instrument plate scale, using the JPL HORIZONS ephemerides.
For each detector pixel the albedo map is integrated over the spatial region that corresponds to the polygon enclosed by the corners of that pixel, creating an albedo array, $\alpha_{x,y}$. If the corner of a pixel is located off the edge of the disk then the region of the albedo map that is enclosed by the remaining three corners is used.   

Since the observed 2.018\,$\mu$m albedo has a contribution from atmospheric scattering, it would be a mistake to use $\alpha_{x,y}$ as the input for the surface reflectivity, $R_{x,y}$, in the model. This would result in an overestimate of the flux at surface-probing wavelengths. We use a linear scaling and offset of the entire map to convert from albedo to reflectivity,

\vspace{-2ex}
\begin{equation}
R_{x,y} = \alpha_{x,y}p_{1}+p_{2}
\end{equation}
where the parameters are determined by fitting the output model spectrum to the observations. A series of models are calculated over a range of input values for $p_{1}$ and $p_{2}$. In each case the mean squared residual between the model and observations is calculated over the entire datacube, and the parameter combination corresponding to the model with the smallest residual is selected as the best fit, $p_{1}$=0.83 and $p_{2}$=0.02

In the radiative transfer calculation we assume isotropic scattering at the surface. As described in \cite{Toon1989}, the total upward flux from the surface is the sum of the reflected downward diffuse flux and flux from any other sources, $S_{\mathrm{sfc}}$. At 2\,$\mu$m, the other relevant source of flux is the reflection of the incident sunlight that reaches the surface, given by

\vspace{-3ex}
\begin{equation}S_{\mathrm{sfc}} = R_{x,y} \mu_{0} \exp{(-\tau/\mu_{0})}\pi F_{s}
\end{equation} 
where $\mu_{0}$ is the cosine of the solar incidence angle, $\tau$ is the optical depth of the atmosphere, and $\pi F_{s}$ is the incident solar flux. 

The Huygens probe measured the surface reflectivity spectrum {\em in situ} from the visible up to 1.5\,$\mu$m \citep{Tomasko2005}. At longer wavelengths, remote observations through narrow regions of low methane opacity are only sensitive to portions of the surface reflectivity spectrum. Since a flat spectrum does not reproduce the observations, i.e. the surface reflectivity in the 1.6\,$\mu$m window is systematically larger than in the 2.0\,$\mu$m window and the contrast between bright and dark regions is different \citep{Coustenis2005,dePater2006},  we incorporate a surface reflectivity spectrum in our model.

We use a set of Gaussian features in the surface spectrum to resolve discrepancies between the calculated albedo and the observations. The surface reflectivity for the model datacube is  given by

\vspace{-2ex}
\begin{equation}
R(x,y,\lambda) = R_{x,y} \prod_{i=0}^{n}\left( 1 - A_{i}\exp ({-(\lambda_{c,i} - \lambda)}/{\delta\lambda_{i}})^{2}\right)
\end{equation}
where we call the product term the surface reflectivity scale factor. Each spectral feature, $i$, is defined by a central wavelength, $\lambda_{c,i}$, line width, $\delta\lambda_{i}$, and amplitude, $A_{i}$, for $n$ features. Values for the Gaussian features are given in Table \ref{tbl2}. Three narrow absorptions are used in the base model and (since $R_{x,y}$ is determined at 2\,$\mu$m) another Gaussian feature is used to represent the increase in reflectivity toward 1.6\,$\mu$m.  It is unclear if these transitions can be uniquely related to the surface composition. The surface reflectivity scale factor is fit globally and individual spectra in the observed datacubes are integrated over vast regions covering more than 10$^{5}$\,km$^{2}$. 

\begin{figure}
\includegraphics[width=5.5in]{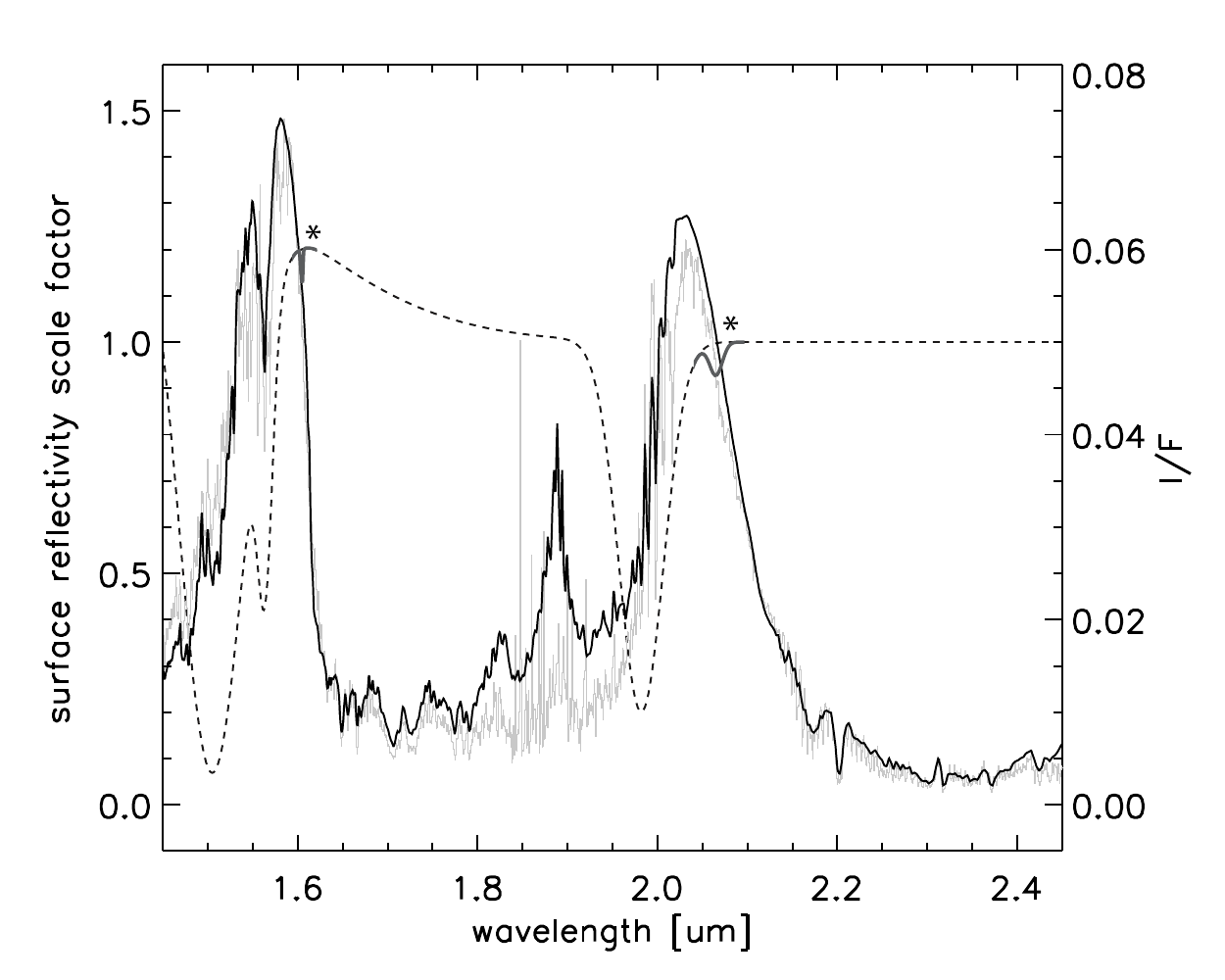}
\caption{ Surface reflectivity spectrum used in RT models is a combination of the surface albedo measured by Cassini/VIMS at 2.018$\,\mu$m and a wavelength-dependent scale factor (dotted line, left axis). The surface reflectivity spectrum is used to calculate the $I/F$ spectrum (thick line, right axis) and compare to observations (thin line, right axis). The base model surface reflectivity scale factor (dotted line, left axis) fits both the dark and bright regions, and a contrived spectrum with two additional absorptions (indicated near `*') is used only for the region of Xanadu to test the differential imaging analysis.\label{fSurfSpec}}
\end{figure}

\begin{figure}
\includegraphics[width=5.25in]{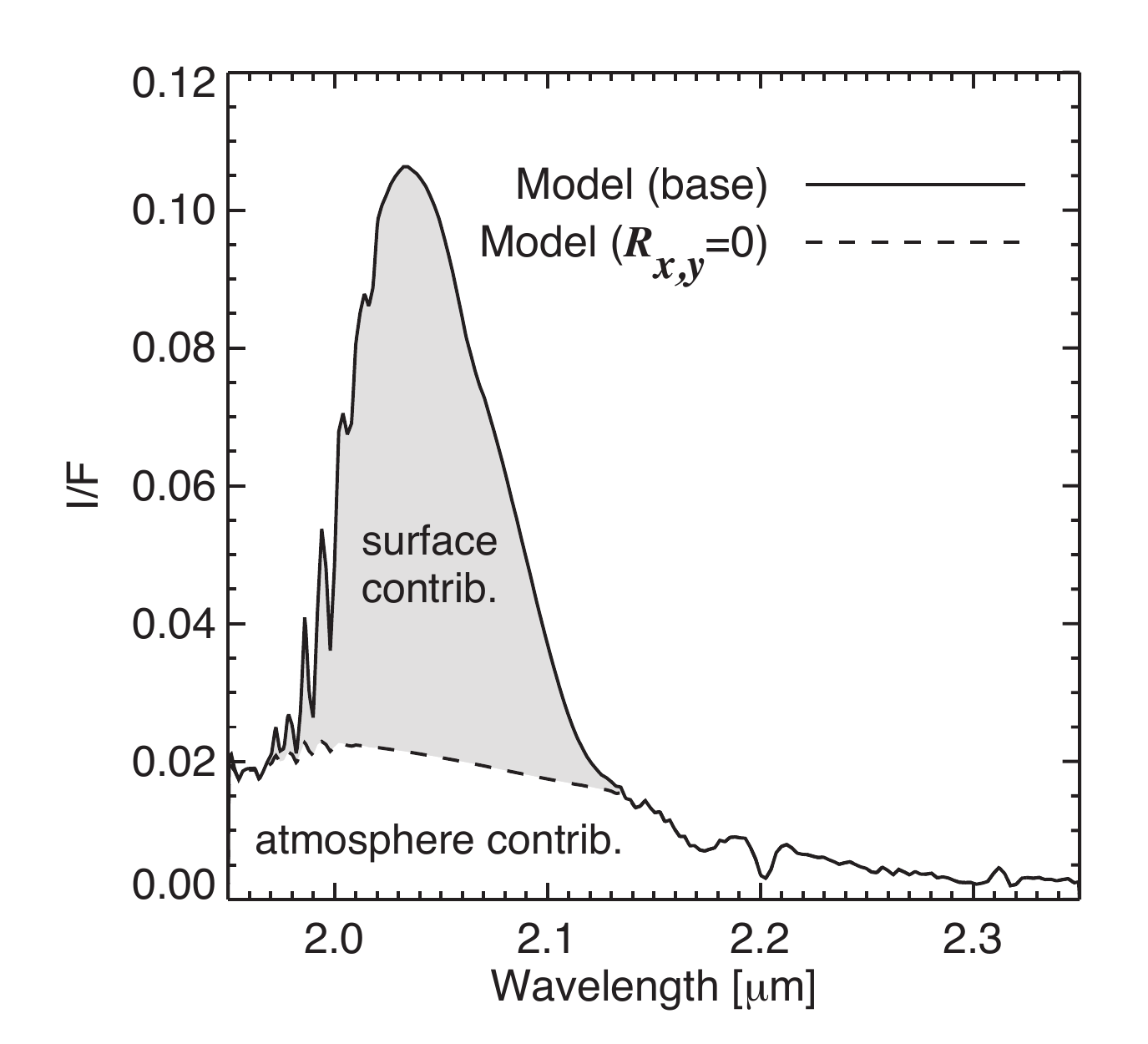}
\caption{The contribution of flux from the surface in the 2\,$\mu$m window was calculated by comparing a spectrum of a bright region near the center of the disk in the base model (solid line) with a model that has the surface reflectivity, $R_{x,y}$, set to zero (dotted line). In the case that $R_{x,y}$=0, all of the flux is from scattering in the atmosphere. The atmospheric scattering is assumed to be the same in the base model, which also includes the light that is reflected from the surface. The maximum surface contribution occurs at 2.04\,$\mu$m, where the flux from the atmosphere is 17\% of the total flux. Integrating the surface contribution over the entire 2\,$\mu$m window gives a minimum atmospheric contribution of 32\% of the total observed albedo. \label{fsurfcontrib}}
\end{figure}

The surface reflectivity scale factor is plotted in Figure~\ref{fSurfSpec} with an example comparison of the model output and observations. By considering the surface reflectivity a free parameter, it is possible to mask uncertainties in atmospheric opacity and fit the spectra arbitrarily well at wavelengths that sense the surface. The relative contribution of surface flux depends on wavelength, surface reflectivity, and viewing geometry. To approximate the fraction of flux arising from the surface in the 2\,$\mu$m window, we compare a spectrum calculated with the base model to a model spectrum with the surface reflectivity, $R_{x,y}$, set to zero; the two spectra are plotted in Figure \ref{fsurfcontrib}. The minimum contribution of the atmospheric flux is 17\% of the total , at 2.04\,$\mu$m for a surface reflectivity of 0.12, and will be larger for lower reflectivity surfaces, and locations near the edge of the disk where the surface is limb-darkened while the atmosphere limb brightens.

\subsection{Condensed-phase methane}

After Voyager it was suggested that methane is saturated in some regions of Titan's atmosphere \citep{Flasar1981, Toon1988}. The methane relative humidity profile measured by Huygens \citep{Fulchignoni2005} and the presence of clouds both conclusively indicated that methane exists in solid and liquid form in Titan's atmosphere. Despite this evidence, RT models used for direct comparison to near-IR observations have only recently included a rudimentary treatment of condensed-phase methane opacity \citep{Adamkovics2007}. The morphology of condensed-phase methane (e.g., particles, droplets, or grains) is unconstrained by these models. A solid cloud of methane \citep{Tokano2006a, Adamkovics2007} is included in the 25-35\,km altitude layer of the model with a methane column volume of 1.5\,cm$^{3}$\,cm$^{-2}$. The column volume is equivalent to a uniform layer of condensed methane with a thickness, $\ell$=1.5\,cm. The optical depth of condensed-phase methane is given by 

\vspace{-2ex}
\begin{equation}
\tau_{c}(\lambda) = \ell \alpha(\lambda)
\end{equation}
where the absorption coefficient spectrum, $\alpha(\lambda)$, is from laboratory measurements \citep{Grundy2002}. The base model has no additional methane precipitation. This basic treatment of the condensed phase opacity assumes only absorption and negligible scattering --- i.e, the scattering at these altitudes is dominated by aerosol haze. 

\subsection{Point spread function}

The surface reflectivity (from the VIMS albedo map) and the stratospheric aerosol gradient (a free parameter) are the two spatially-varying inputs in the model. After these inputs have been determined, the RT calculation of a spectrum is performed for each of the approximately 4000 spatial locations, building a model datacube at the plate scale of the observations. The last step is to convolve the model with a point spread function (PSF). We use the telluric calibration star datacube, collapsed in wavelength across H- and K-bands separately, as the reference images for the PSF. A 2d Gaussian is fit to the PSFs and used as the convolution kernel for each band. Fitting the PSF and then using centered convolution kernel avoids the addition of noise into the model. 

\section{Surface Subtraction}\label{sDI}

Atmospheric scattering at 2\,$\mu$m accounts for at least $\sim$17\% of the total observed flux, the rest is from the surface. Changes in surface reflectivity at different locations on Titan generally swamp the signal from spatial variations in atmospheric scattering. 

Images probing the lower atmosphere are also sensitive to the surface. The latter can be subtracted as described below, to discern atmospheric phenomena \citep{Adamkovics2007}.  
We calculate surface subtracted, $\Delta I/F$, images using the method of \citet{Adamkovics2007} for all of the observations and corresponding models. Each pixel in the subtracted images is defined according to

\begin{figure}\begin{center}
		\includegraphics[width=5.25in]{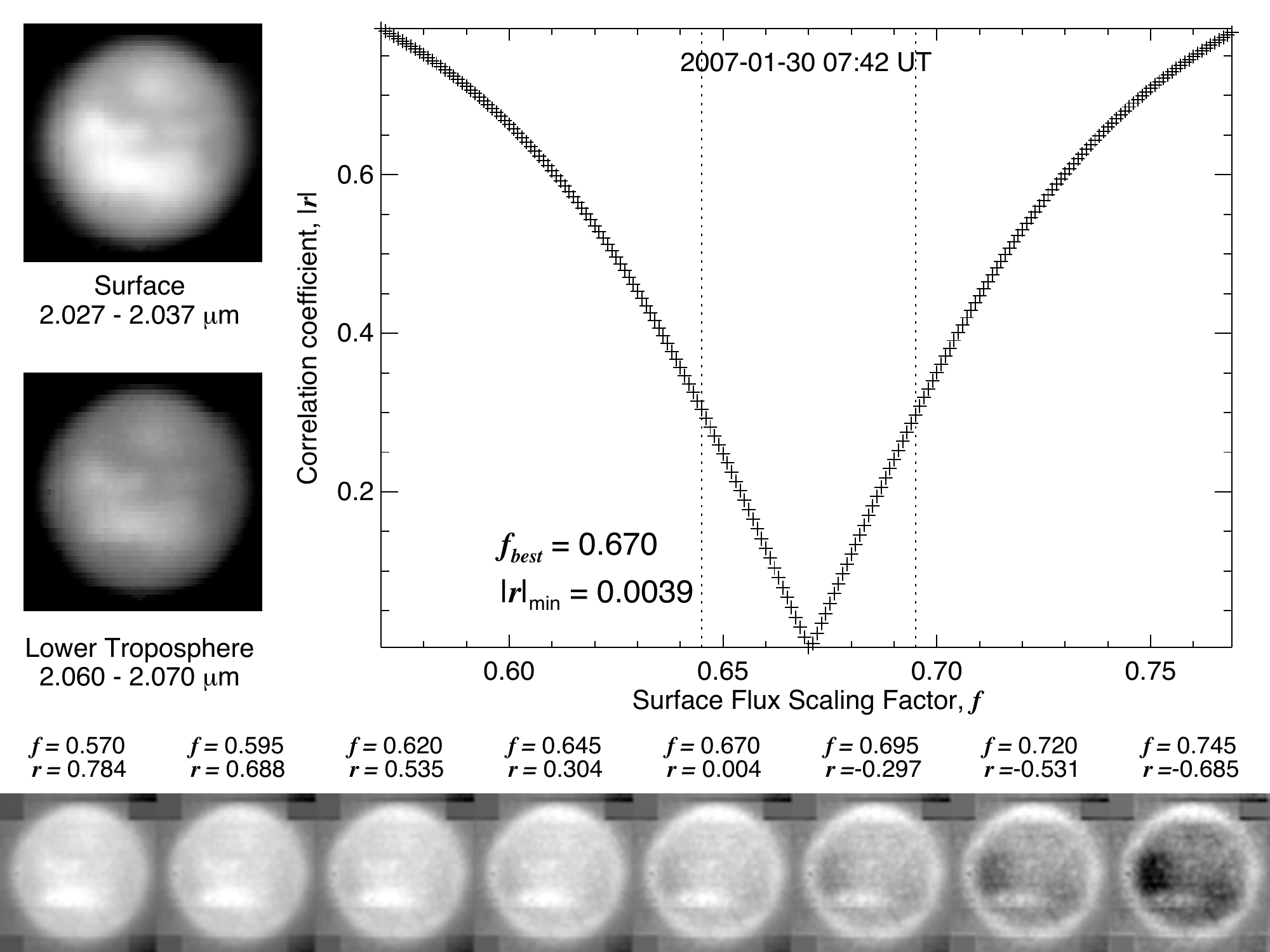}
        \end{center}        
   \caption{The scaling factor for the image of the surface flux is determined by an empirical test of the correlation between the surface probing image and the subtracted image.}
\label{fsftest}	
\end{figure}

\vspace{-3ex}
\begin{equation}
   \Delta I/F = (I/F)_{\lambda_{1}} - f \times (I/F)_{\lambda_{2}} ,
\end{equation}
where the bandpasses for $(I/F)_{\lambda_1}$ and $(I/F)_{\lambda_2}$ are given in Table \ref{tdibp}. The bandpass for $(I/F)_{\lambda_2}$ is selected to cover the wavelength region that probes the surface with the least atmospheric contribution, while $(I/F)_{\lambda_1}$ is integrated over a bandpass with increased methane absorption and therefore contains a larger contribution from the atmosphere.  The linear Pearson correlation coefficient, $r$, between the $\Delta I/F$ image and the $(I/F)_{\lambda_{2}}$ is determined for a range of $f$. Surface subtracted images using a range of values for $f$ demonstrate that small values ($f < 0.57$) yield images correlated with surface probing images, while large values ($f > 0.74$) are anti-correlated with the surface image --- that is, they yield and inverted image of the surface albedo pattern and large negative $r$ (Figure~\ref{fsftest}). Intermediate values minimize the absolute value of the correlation coefficient.
Although the contribution of flux from the surface depends on a reflectivity, the value of $f$ that corresponds to minimum $|r|$ is the optimal value used in this simplistic image subtraction. A single $f$ may be used in a model-independent, approximate method for removing surface albedo variation from images.  

Since the background noise in regions of the image with sky can affect the statistical analysis, we performed the calculation of $f$ for each night using only the pixels on the disk of Titan, yielding $\bar{f}$=0.67, with  a mean absolute deviation $\sigma$=0.011. The scale factors for each night are presented in Table \ref{tsfpar} for the 1.5\,$\mu$m and 2\,$\mu$m bands. These values are significantly lower than when the analysis is performed on the entire image, which includes noise from the sky; in that case $\bar{f} $= 0.76 and $\sigma$=0.013. The values of $f$ calculated while including regions of the sky are consistent with the smallest value calculated by \citet{Kim2008} and larger than  $f$=0.72 used in \citet{Adamkovics2007}.  The $\Delta I/F$ images for the observations and all models --- each calculated with $\bar{f}$=0.67 --- are presented in the bottom four rows of Figure~\ref{fObs}. 

Comparison of the observations and the models, in both wavelength regions ($\lambda_{1}$ and $\lambda_{2}$) demonstrates that where the surface albedo is well mapped by Cassini/VIMS, the models reproduce the observations well. Additionally, changes due to additional condensed-phase methane opacity do not produce significant changes to the images in the $\lambda_{2}$ bandpass. The small variations in the flux from subtle changes in opacity and reflectivity are more clearly observed after subtraction of the surface albedo variation. The 1$\sigma$ pixel-to-pixel noise is between 0.0012 and 0.0016 for all of the observed $\Delta I/F$ images. These values are calculated two ways: first, a conservative estimate is made by calculating the mean of the absolute value of the residuals between the base model and the observations, and secondly by calculating the FWHM of the histogram of the $\Delta I/F$ values at the center of the disk in each image. In the following discussion, we investigate the suggestion of \citet{Kim2008} that bright regions of the surface cause artifacts in the difference images.

\section{Discussion}

\begin{figure*}
\vspace{-2cm}
\includegraphics[width=5.35in]{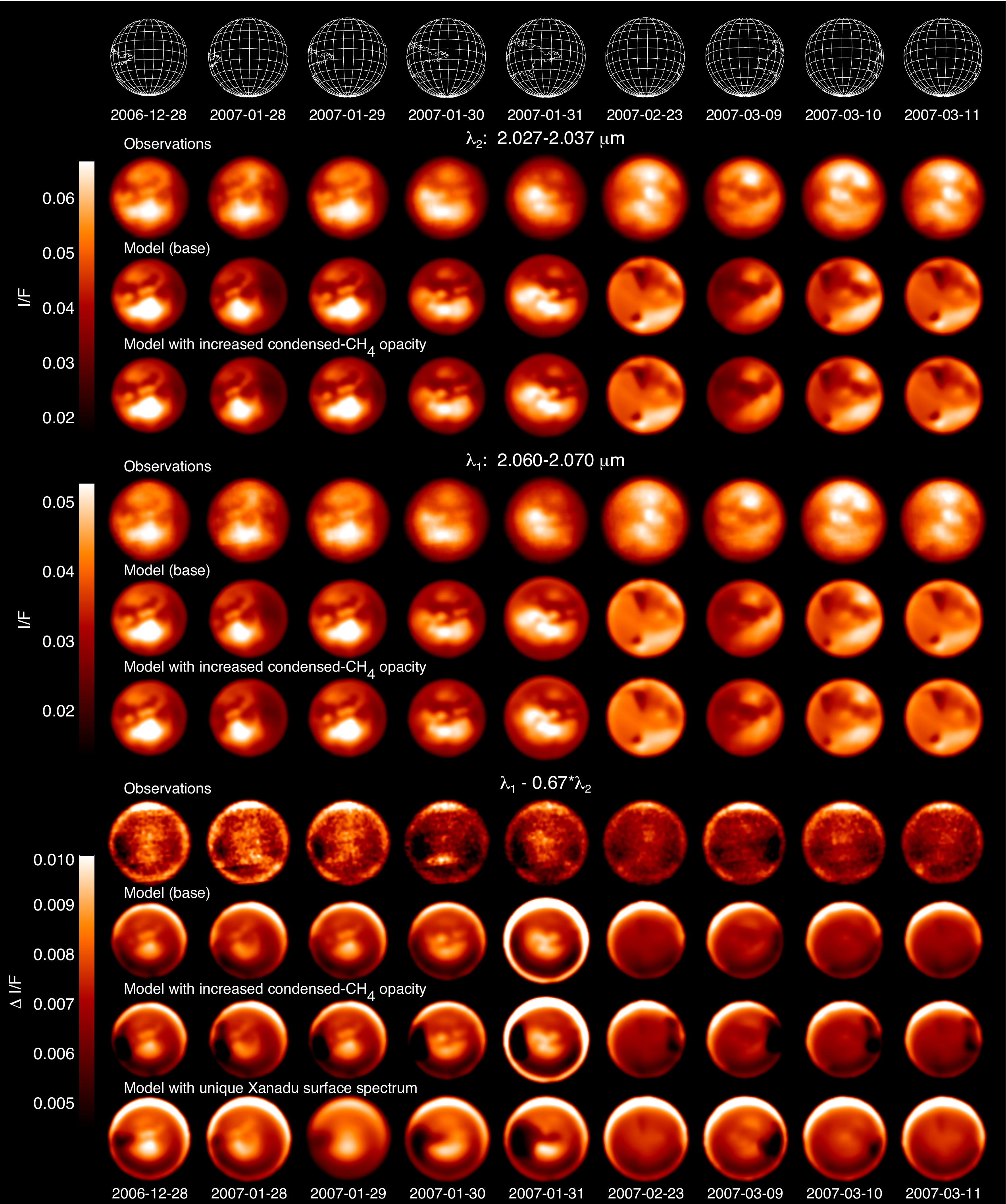}
\caption{ K-band (2\,$\mu$m) VLT/SINFONI observations and radiative transfer models both with and without an increase in condensed phase methane near Xanadu. The increase is modeled to occur from 30$^{\circ}$S to 20$^{\circ}$N, and from the limb to within 35$^{\circ}$ of the sub-observer longitude for all observations. A prominent southern mid-latitude cloud (identified at longer wavelengths) contaminates the $\Delta I/F$ image on 2007-01-30 UT, appearing as a bright feature between 30 -- 45$^{\circ}$S. An incorrectly high value for $f$ affects areas with the same surface albedo uniformly and would result in all of Titan's bright regions showing small $\Delta I/F$ values. This is not the case here. While Xanadu corresponds to low $\Delta I/F$, other high-albedo surface terrains such as Dilmun (10$^{\circ}$N 175$^{\circ}$W), Tsegihi (40$^{\circ}$S, 35$^{\circ}$W), and Adiri (10$^{\circ}$S 210$^{\circ}$W) do not show similar negative excursions --- however, this is not the case for Figure~3b in \citet{Kim2008}. \label{fObs}}
\end{figure*}

The ground-based observations and RT models are presented in Figure~\ref{fObs}.  Although the entire observed datacube is calculated (covering 1.45 -- 2.45\,$\mu$m) we focus here on images from the wavelength windows that are used to identify condensed phase methane opacity.

\citet{Kim2008} argue that anti-correlations are created in $\Delta I/F$ images due to regions of high surface albedo having different $f$ values than dark regions. The argument predicts that bright regions will, in general, be anti-correlated with dark regions in $\Delta I/F$ images. Observations in Figure~\ref{fObs} demonstrate that a gross anti-correlation does not occur. The brightest regions in the top row of images does not correspond with the dark regions in the subtracted images. Due to limb darkening, regions of the surface with the brightest $(I/F)_{\lambda_{2}}$ always occur at the center of the disk, whereas dark regions in the $\Delta I/F$ (if present at all) occur near the limb. There is no gross anti-correlation between $(I/F)_{\lambda_{2}}$ and $\Delta I/F$ images.

Perhaps the argument of \citet{Kim2008} applies only to the region of highest surface reflectivity, Xanadu, or as they suggest, the spectrum of the surface needs to be taken into consideration. The four consecutive nights of observations from 2007-01-28UT through 2007-01-31UT are used to test this hypothesis. On 2007-01-28UT the eastern tip of Xanadu is visible on the limb of Titan, and by 2007-01-31UT this location is observed at the center of the disk. In the $\Delta I/F$ images from these dates, there is a darker region over the tip of Xanadu on 2007-01-28UT, however, the same location is no longer dark on 2007-01-31UT. This sequence of observations suggest that artifacts related to static properties of the surface --- which are expected to track with rotation --- do not account for the dark regions of the surface. To further test this hypothesis, we employed a systematic change in the surface reflectivity spectrum, but only near Xanadu. 

The Xanadu region is outlined (somewhat arbitrarily) using a high resolution ISS map \citep{Porco2005}. A test case is modeled where the spatial locations in the observed datacubes that are bounded by Xanadu (plotted in the top row of Figure~\ref{fObs}) have two additional surface absorption features, listed in Table \ref{tbl2}, and shown graphically in Figure~\ref{fSurfSpec}. The features are centered on the wavelength regions that are used to probe the atmosphere, above the surface in both H- and K-bands, and are narrow enough to fit within the $\lambda_{2}$ bandpass used for the images. The effect of this mildly contrived change on the surface-subtracted images is demonstrated in the bottom row of Figure~\ref{fObs}. Not unexpectedly, the region of Xanadu is darker in the difference images. Altering the surface spectrum in such a way could be modified slightly to fit an individual observation very well. However, the series of observations from 2007-01-28 through 2007-01-31 illustrate that a systematic change to the surface reflectivity over Xanadu cannot reproduce all the observations. For these dates, the  mean of the absolute value of 
residuals between the condensed-phase model and the observations is smaller than the residuals of the model with the artificial surface spectrum. On 2007-02-23, 2007-03-10, and 2007-03-11, the models with either additional methane opacity or the surface spectrum change both have larger residuals than the base model.

\begin{figure}
\begin{center}
\includegraphics[width=5.25in]{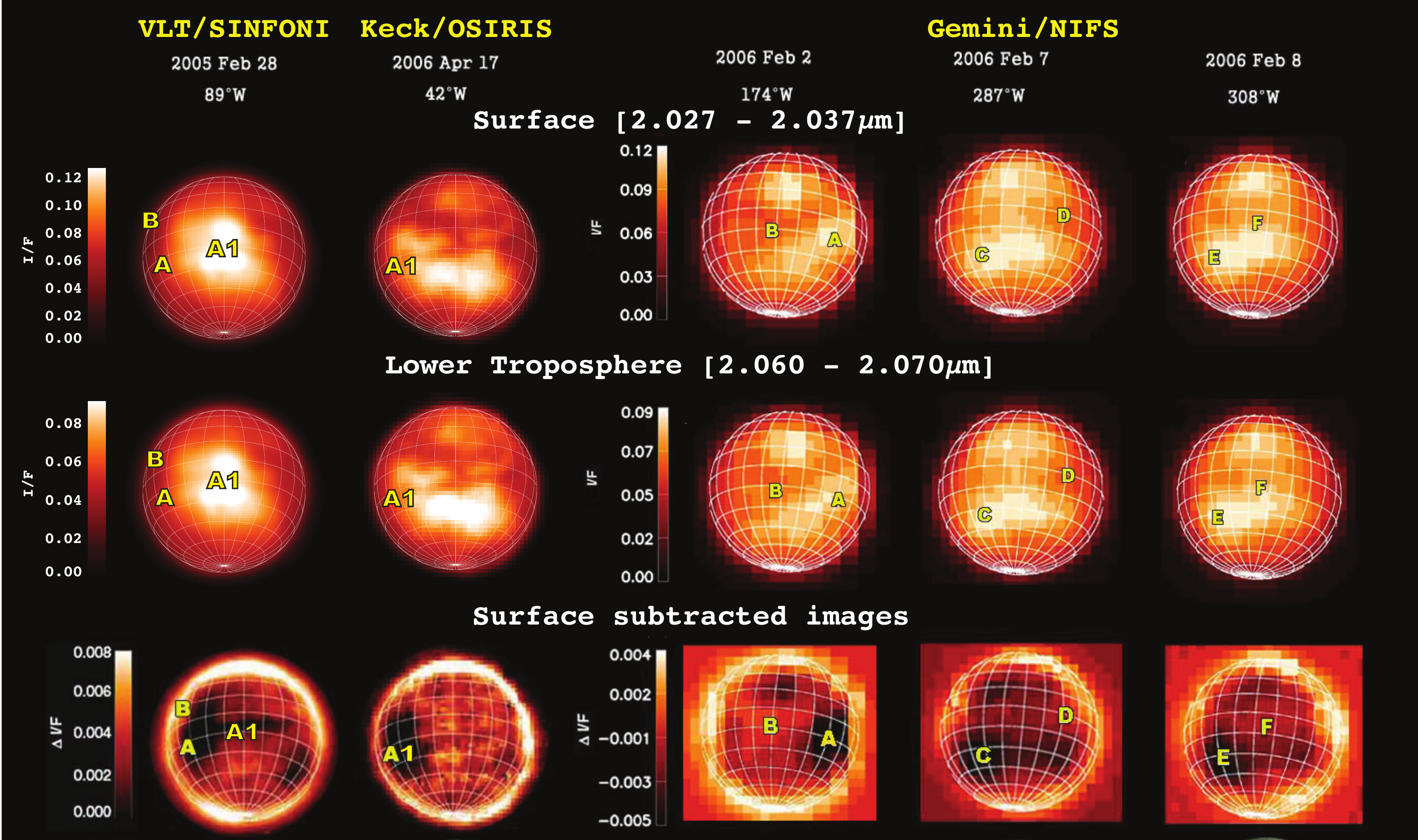}
\caption{Comparison of field-integral spectrometer images from three instruments: VLT/SINFONI, Keck/OSIRIS, and Gemini/NIFS. The VLT and Keck data are from \citet{Adamkovics2007}, while with lower S/N and lower resolution Gemini data are combined from Figures 1 and 3 in \citet{Kim2008}. For clarity, the two regions labeled in `A' by \citet{Kim2008} are separated here as `A1' and `A'. The `A1' regions are dark in the difference image from Keck
while appearing bright in the VLT observations. \label{fGVK}}
\end{center}
\end{figure}

The observed $\Delta I/F$ image on 2007-01-29 has a darker region near the limb than in the model with a unique Xanadu surface spectrum. This suggests that the absorption spectrum of the Eastern tip of Xanadu is underestimated in the model. However, on 2007-01-31 this same region of Xanadu is near the center of the disk and the absorption spectrum is overestimated in the model. A change to the surface reflectivity spectrum cannot simultaneously match both observations. A similar effect was noted qualitatively in \citet{Adamkovics2007}, and is also illustrated in Figure~\ref{fGVK} with the observations from \citet{Kim2008}. Whereas the two regions of Eastern and Western Xanadu are both labelled `A' in Figures 1 and 3 by \citet{Kim2008}, here we label the East and West regions separately as `A' and `A1', respectively. This clarification shows that while `A1' is a bright albedo region that shows up as a dark feature in the Keck/OSIRIS difference images from 2006 Apr 17, this same bright feature --- near the center of disk --- is no longer a dark region in the difference image viewed with VLT/SINFONI.

\begin{figure}
\includegraphics[width=5.25in]{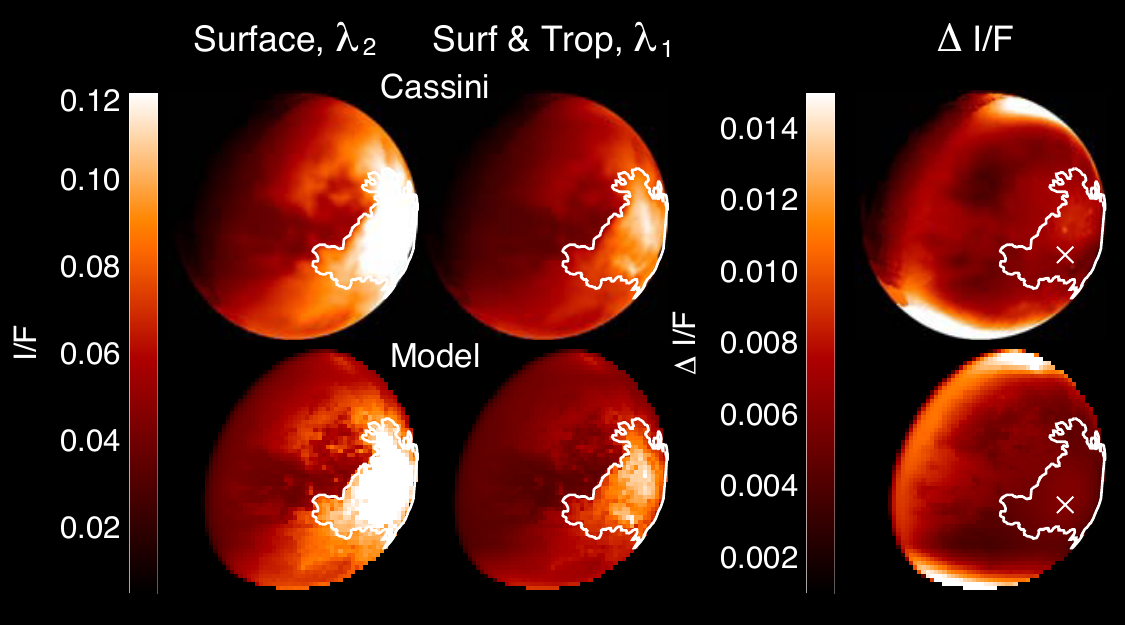}
\caption{Cassini/VIMS observations and surface subtracted images at high phase angle (top), with radiative transfer models of the same observing geometry (bottom). The outline indicates Xanadu. Limb brightening in the $\Delta I/F$ images due aerosol scattering is reproduced in the models. Bright scattering at the poles is due to optically thin clouds.\label{fCass}}
\end{figure}

We test for surface scattering phase artifacts by performing our analysis on Cassini/VIMS data from the T10 flyby. 
Since the dark regions in the $\Delta I/F$ images occur near the limb, it is possible that some property of the surface scattering is the cause of an artifact that causes Xanadu to always be dark in the $\Delta I/F$ images near the limb.
However, the images in Figure~\ref{fCass} demonstrate that Xanadu is not an inherently dark region in $\Delta I/F$ images. Intensity variation in the VIMS $\Delta I/F$ image does indicate that a single value of $f$ will not remove all the spatial variation in the subtracted image, however, the gross over-correction for bright regions is avoided for a minimum value of the correlation coefficient, $|r|$. Our RT models also reproduce the high phase angle observations with Lambertian surface scattering and a surface reflectivity scale factor spectrum that does not vary across the disk. 

\begin{figure}
\includegraphics[width=5.25in]{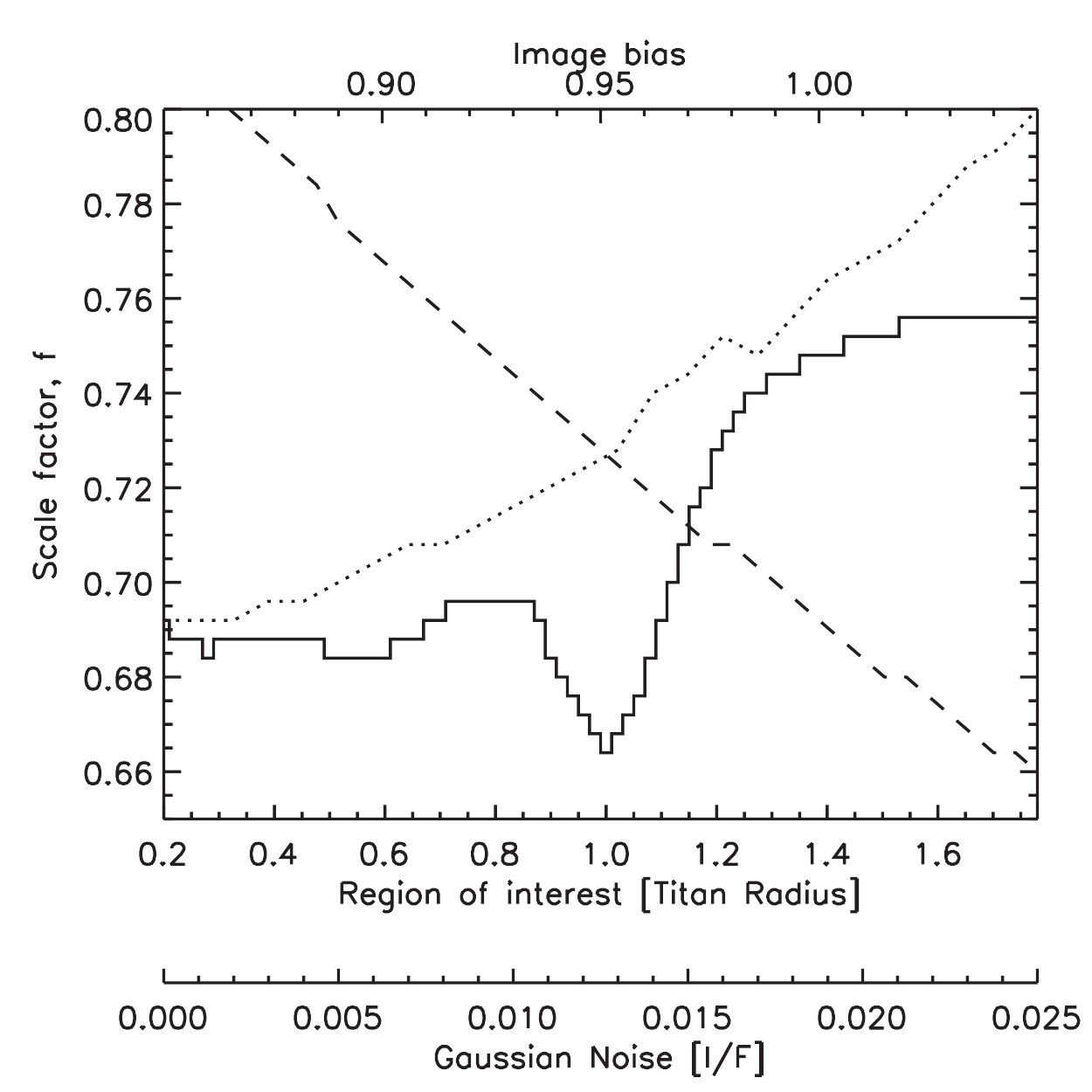}
\caption{The optimal surface-subtraction scale factor $f$  depends on the region of interest in the image (solid line, lower axis), bias between the $(I/F)_{\lambda_1}$  and $(I/F)_{\lambda_2}$ image (dashed line, top axis), and the noise in the analyzed images (dotted line, lowest axis). Calculations for bias and noise were made using $R_T<0.7$. \label{fnoise}}
\end{figure}

The observations and models presented here can be reconciled with the measurements of \citet{Kim2008} by quantitatively evaluating factors that can affect the determination of $f$.  We calculate $f$ for a subset of pixels in a circle centered on Titan and plot the resultant $f$ as a function of Titan radius, $R_T$, in Figure~\ref{fnoise}). Near the center of the disk  ($R_T<0.7$) , $f$ is roughly constant. There is a decrease in $f$ when limb-darkened regions ---  where contributions from the surface are small --- are included in the correlation analysis. Including regions of the sky adds noise and increases the calculated value of $f$.  We further investigate this effect by systematically adding noise to the images and reproducing our analysis. We generate images with Gaussian noise (of a given FWHM in $I/F$), add the image of noise to our observations, and then recalculate $f$ (lower scale bar in Figure~\ref{fnoise}). Additional noise leads to a systematically higher determination of $f$ (dotted line Figure~\ref{fnoise}) such that 1\% of additional noise can lead to a increase in $f$ by $\sim$0.02. A linear bias is similarly tested by scaling the $(I/F)_{\lambda_{2}}$ image (dashed line in 
Figure~\ref{fnoise}). This is an approximation of uncertainties due to flat-fielding, bias-correction, or telluric correction. Both positive or negative deviations in $f$ can occur due to biased images. Data from 2007-01-28UT are used above for illustration, however the results generalize to each night of our observations. These tests demonstrate that using the entire image in the analysis, an offset during data reduction, or a noisier dataset can all lead to larger values of $f$, and thus an `over-subtraction' of the surface. Overestimating $f$  will lead to anti-correlated artifacts from regions of bright surface reflectivity, as detailed in \citet{Adamkovics2007}, and evidenced in \citet{Kim2008}.  Figure~\ref{fGVK} demonstrates that the Gemini/NIFS data presented by \citet{Kim2008} are noisier than either the VLT/SINFONI or Keck/OSIRIS images presented in \citet{Adamkovics2007}.

\section{Conclusions}

Atmospheric phenomena are most likely responsible for the observed dark regions in the $\Delta I/F$ images and have been quantitatively reproduced in models using additional condensed-phase methane (as in  \citet{Adamkovics2007}) and tested here against surface scattering artifacts. Since the extinction properties of haze are not strongly wavelength-dependent in the 2\,$\mu$m region, spatial variations in haze are removed during the image subtraction and cannot reproduce the dark regions in $\Delta I/F$ images. Altering the haze properties enough to create the dark  $\Delta I/F$  regions results in discrepancies that fail to reproduce the observed albedo in tropospheric or stratospheric images \citep{Adamkovics2007}.
Localized meteorology has already been observed \citep{Roe2005b}, so perhaps precipitation is also geographically linked. On 2007-03-09 we observe condensed-phase methane over Xanadu on the evening limb, Figure~\ref{fObs}, indicating that precipitation can occur either in the morning or in the afternoon.

Further modeling of scattering by strongly absorbing droplets or ices is needed to determine if the rain can reach the ground and if there is a predicted signature of rain at longer wavelengths (e.g, 5\,$\mu$m). There is also a possibility that the condensed methane coexists as a surface coating on aerosols. Alternatively, the optical properties of haze may be such that they mimic methane transmission spectrum,  which is unexpected but cannot be ruled out. Currently, enhancement in condensed-phase methane over Xanadu remains the least speculative and the only modeled description of dark regions in $\Delta I/F$ images.

\section*{Acknowledgments}

This work was supported by NSF and the Technology Center for Adaptive Optics, managed by the University of California at Santa Cruz under cooperative agreement AST-9876783, NASA grant NNG05GH63G, and by the Center for Integrative Planetary Science at the University of California, Berkeley. Observations were performed at the VLT operated by the European Southern Observatory.

\bibliographystyle{elsarticle-harv}

\clearpage

\begin{table*}
\begin{center}
    \small
\caption{Observation dates and viewing geometry.}
	\begin{tabular}{lccccccc}
	\hline
	\hline
		Obs. Date   & Airmass & Diameter & \multicolumn{2}{c}{Sub-observer Point} & \multicolumn{2}{c}{Sub-solar Point} & Phase Angle \\ 
		  (UT)      &         & (arcsec) & $^{\circ}$W Long. & Lat. & $^{\circ}$W Long. & Lat.     & (degrees)   \\ \hline
		2006 Dec 28 & 1.31    & 0.835    & 22.5         & -12.3    & 26.8      & -14.1    & 4.6$^{*}$    \\ 
		2007 Jan 28 & 1.30    & 0.862    & 1.6          & -13.2    & 3.1       & -13.7    & 1.6$^{*}$    \\ 
		2007 Jan 29 & 1.30    & 0.863    & 23.9         & -13.2    & 25.3      & -13.7    & 1.5$^{*}$    \\ 
		2007 Jan 30 & 1.52    & 0.863    & 48.5         & -13.2    & 49.8      & -13.7    & 1.3$^{*}$    \\ 
		2007 Jan 31 & 1.35    & 0.864    & 68.3         & -13.3    & 69.5      & -13.7    & 1.3$^{*}$    \\ 
		2007 Feb 23 & 1.35    & 0.864    & 228.0        & -14.1    & 226.6     & -13.4    &  1.4 $^{ }$  \\ 
		2007 Mar 09 & 1.41    & 0.855    & 186.0        & -14.5    & 183.3     & -13.2    &  3.0 $^{ }$  \\ 
		2007 Mar 10 & 1.37    & 0.854    & 206.5        & -14.6    & 203.7     & -13.2    &  3.0 $^{ }$  \\ 
		2007 Mar 11 & 1.38    & 0.853    & 229.0        & -14.6    & 226.1     & -13.1    &  3.1 $^{ }$   \\
		\hline  
        Cassini T10 & ---     &  ---     & 173.4        & 0.0      & 127.8     & -18.7    & 48.5        \\
		\hline  

	\end{tabular}
	
\footnotesize{$^{*}$Sub-solar longitude is greater than sub-observer longitude. For these observations, the 
\\ sub-solar point is to the left of the center of the disk in the images in Figure~\ref{fObs}.}

\end{center}

\label{tbl1}
\end{table*}

\begin{table}
\begin{center}

\caption{Surface reflectivity spectrum features}

\label{tbl2}
\small
 \begin{tabular}{cccl}
 \hline
 \hline
  Line center, & Line width, & Amplitude, & Notes \\
  $\lambda_{c,i}$	& $\delta\lambda_{i}$ & $A_{i}$ & \\ \hline
    1.400   	& 0.250 & -0.45$^{*}$ &  base model \\
    1.505	& 0.052 &  0.95   &  base model \\
    1.563	& 0.012 &  0.55   &  base model \\ 
  	1.982	& 0.035 &  0.80   &  base model \\ 
    1.605   	& 0.002 &  0.07   &  only near Xanadu \\    
    2.065   	& 0.010 &  0.07   &  only near Xanadu \\
		\hline  
	\end{tabular}

\footnotesize{$^{*}$ Since $R_{x,y}$ in Equation 3 corresponds to observations at 2\,$\mu$m, \\
                     a negative amplitude is used to represent the increase in reflectivity  \\
                     toward 1.6\,$\mu$m (Figure \ref{fSurfSpec}).}

\end{center}
\end{table}

\begin{table}
\caption{Bandpasses ($\mu$m) used for $\Delta I/F$  images}
\label{tdibp}
  \small
\begin{center}
	\begin{tabular}{lcc|c}
	\hline
	\hline
	                         & \multicolumn{2}{c|}{VLT/SINFONI}     & Cassini/VIMS$^{a}$  \\
			    				& 1.5\,$\mu$m band & 2.0\,$\mu$m band & 	2.0\,$\mu$m band \\ \hline
		Surface				& 1.593 -- 1.596	   & 2.027 -- 2.037 	 &	2.026 -- 2.042   \\
		Surf \& Trop			& 1.603 -- 1.606   & 2.060 -- 2.070  	 &	2.059 -- 2.076   \\
        \hline
	\end{tabular}
	
\footnotesize{$^{a}$Approximated bandpass for individual  VIMS channels centered at \\ 2.034, 2.068, 2.117, and 2.150\,$\mu$m.}
\end{center}
\end{table}

\begin{table}
\begin{center}
	\caption{Surface subtraction parameters.}
	\label{tsfpar}
	\begin{tabular}{lcccc}
	\hline
	\hline
		Obs. Date (UT)	& \multicolumn{2}{c}{1.5\,$\mu$m band} & \multicolumn{2}{c}{2.0\,$\mu$m band} \\
		             	& $f_{best}$ & $|r|_{min}$ & $f_{best}$ & $|r|_{min}$\\ \hline
		2006 Dec 28		& 0.529 & 0.0019 & 0.659      & 0.0023      \\ 
		2007 Jan 28 		& 0.550 & 0.0037 & 0.667      & 0.0057      \\ 
		2007 Jan 29 		& 0.549 & 0.0052 & 0.666      & 0.0014      \\ 
		2007 Jan 30 		& 0.577 & 0.0010 & 0.670      & 0.0038      \\ 
		2007 Jan 31 		& 0.581 & 0.0018 & 0.684      & 0.0058      \\ 
		2007 Feb 23 		& 0.621 & 0.0024 & 0.691      & 0.0027      \\ 
		2007 Mar 09 		& 0.583 & 0.0024 & 0.655      & 0.0007      \\ 
		2007 Mar 10 		& 0.606 & 0.0035 & 0.686      & 0.0027      \\ 
		2007 Mar 11 		& 0.606 & 0.0046 & 0.686      & 0.0040      \\ \hline  
		$\bar{f}_{best}$ & 0.578 &  & 0.674 &                     \\ 
		\hline        
	\end{tabular}
\end{center}
\end{table}

\clearpage

\clearpage

\clearpage

\clearpage

\thispagestyle{empty}

\clearpage

\clearpage

\clearpage

\end{document}